\pdfoutput=1
\documentclass[journal=jpcafh,manuscript=article]{achemso}
\usepackage{natbib} 
\usepackage{achemso} 
\SectionNumbersOn
\usepackage[version=3]{mhchem} % Formula subscripts using \ce{}
\usepackage[T1]{fontenc}       % Use modern font encodings
\usepackage{xcolor}

\usepackage{amsmath}
\usepackage{fp} % floating point calculations

\newcommand{\rndq}[1]{%
\FPeval\val{round(( #1):4)}%
\FPprint{\val}%
}

\newcommand{\rndd}[1]{%
\FPeval\val{round(( #1):2)}%
\FPprint{\val}%
}

\newcommand{\angstrom}{\text{\normalfont\AA}}
\author{Sergio Rampino}
\email{sergio.rampino@sns.it}
\affiliation[SNS]
{Scuola Normale Superiore,
Piazza dei Cavalieri 7,
56126 Pisa, Italia}
\author{Yury V. Suleimanov}
\email{y.suleymanov@cyi.ac.cy, ysuleyma@mit.edu}
\affiliation[CYI]{Computation-based Science and Technology Research Center, Cyprus 
Institute, 20 
Kavafi Street, Nicosia 2121, Cyprus}
\alsoaffiliation[MIT]{Department of Chemical Engineering, Massachusetts Institute of Technology, 
Cambridge, Massachusetts 02139, United States}

\title[Short title]%
{Thermal
Rate Coefficients
for the Astrochemical Process C + CH$^+$ $\to$ C$_2^+$ + H by Ring Polymer Molecular Dynamics%
}

\begin{document}
%%%%%%%%%%%%%%%%%%%%%%%%%%%%%%%%%%%%%%%%%%%%%%%%%%%%%%%%%%%%%%%%%%%%%
%% The "tocentry" environment can be used to create an entry for the
%% graphical table of contents. It is given here as some journals
%% require that it is printed as part of the abstract page. It will
%% be automatically moved as appropriate.
%%%%%%%%%%%%%%%%%%%%%%%%%%%%%%%%%%%%%%%%%%%%%%%%%%%%%%%%%%%%%%%%%%%%%
\begin{tocentry}
%
% srampino chk fill toc
%\vskip -0.9cm
%\includegraphics[width=2.0in]{./pdf/toc-crop.pdf}
\includegraphics[height=3.5cm]{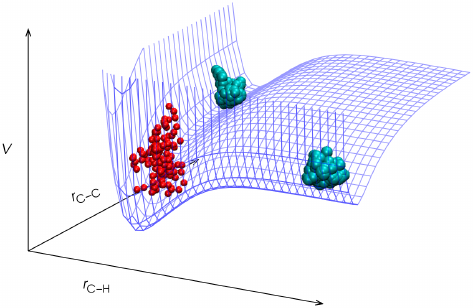}
% end srampino
%Some journals require a graphical entry for the Table of Contents.
%This should be laid out ``print ready'' so that the sizing of the
%text is correct.
%
%Inside the \texttt{tocentry} environment, the font used is Helvetica
%8\,pt, as required by \emph{Journal of the American Chemical
%Society}.
%
%The surrounding frame is 9\,cm by 3.5\,cm, which is the maximum
%permitted for  \emph{Journal of the American Chemical Society}
%graphical table of content entries. The box will not resize if the
%content is too big: instead it will overflow the edge of the box.
%
%This box and the associated title will always be printed on a
%separate page at the end of the document.

\end{tocentry}

%%%%%%%%%%%%%%%%%%%%%%%%%%%%%%%%%%%%%%%%%%%%%%%%%%%%%%%%%%%%%%%%%%%%%
%% The abstract environment will automatically gobble the contents
%% if an abstract is not used by the target journal.
%%%%%%%%%%%%%%%%%%%%%%%%%%%%%%%%%%%%%%%%%%%%%%%%%%%%%%%%%%%%%%%%%%%%%
% srampino
\newpage
% end srampino
\begin{abstract}

Thermal rate coefficients for the astrochemical reaction C + CH$^+$ $\to$ C$_2^+$ + H were computed in the temperature range 20-300 K by using novel rate theory based on ring polymer molecular dynamics (RPMD) on a recently published bond-order based potential energy surface and compared with previous Langevin capture model (LCM) and
quasi-classical trajectory (QCT) calculations. Results show that there is a significant discrepancy between the RPMD rate coefficients and the previous theoretical results which can lead to overestimation of the rate coefficients for the title reaction by several orders of magnitude at very low temperatures. We argue that this can be attributed to a very challenging energy profile along the reaction coordinate for the title reaction, not taken into account in extenso by either the LCM or QCT approximation. In the absence of any rigorous quantum mechanical  or  experimental results, the computed RPMD rate coefficients represent state-of-the-art estimates to
be included in astrochemical databases and kinetic networks.

%The temperature dependence of the computed rate coefficients is analyzed in relation to the Arrhenius-Kooij and Aquilanti-Mundim laws.
%This is also the longest real-time dynamics (16 ps) ever studied with RPMD and it is a real complex-formation reaction.

\end{abstract}

%%%%%%%%%%%%%%%%%%%%%%%%%%%%%%%%%%%%%%%%%%%%%%%%%%%%%%%%%%%%%%%%%%%%%
%% Start the main part of the manuscript here.
%%%%%%%%%%%%%%%%%%%%%%%%%%%%%%%%%%%%%%%%%%%%%%%%%%%%%%%%%%%%%%%%%%%%%
% srampino
\newpage
% end srampino
\section{Introduction}

One of the main goals of astrochemistry is to develop a complete chemical model 
of interstellar clouds giving an account of the nature and
abundance of the molecules observed in the interstellar medium (ISM).
This involves solving a large number of rate equations describing the changes
in the concentration of chemical species as a result of processes---mainly
gas-phase collisions or processes occurring on the surface of ice or dust
 particles \cite{herbst01_168}---where hundreds of species act simultaneously
as reactants and products.
%Present models account for
%several hundreds of species connected by thousands of reactions
%(468 species and 6046 reactions, for instance, in the OSU\_01\_2009
%gas-phase
%model) \bibnote{See \url{http://faculty.virginia.edu/ericherb/research.html}, accessed 29 August 2016}.
Clearly, it is crucial that the main properties of these processes, like the
chemical rate coefficients, are estimated as accurately as possible.

Chemical rate coefficients for inclusion in kinetic models are made available
through online kinetic databases such as the Ohio State University
OSU database
(available at \url{http://faculty.virginia.edu/ericherb/research.html}),
the KInetic Database for Astrochemistry KIDA \cite{wakelam12_21}
(available at \url{http://kida.obs.u-bordeaux1.fr/}),
and the University of Manchester Institute of Science and Technology (UMIST)
Database for Astrochemistry UDfA \cite{woodall07_1197,mcelroy13_a36}
(available at \url{http://udfa.ajmarkwick.net/}).
However, many of the estimates for the rates of the reactions presently
available in these
databases lack a sound foundation (being sometimes based on simplified models,
or calculated by analogy with similar systems where the rate constant 
is known, or worked out by extrapolation to low temperature from
high-temperature estimates)
and a revision, possibly based on rigorous quantum dynamics techniques, is
often in order.

The title reaction, belonging to the important class of astrochemical processes 
involving carbon and hydrogen atoms \cite{solomon72_389,herbst01_168}, is among
those reactions whose dynamics and kinetics have been little investigated.
The lowest-energy reactive channel for collision of C + CH$^+$ is
\begin{equation}\label{eq:reaction}
\text{C ($^{3}P_0$) + CH$^+$ ($X{}^1\Sigma^+$) $\to$ C$_2^+$ ($X{}^4\Sigma^-_\mathrm{g}$) + H ($^{2}S_{1/2}$)} %\;,
%\text{C + CH$^+$ $\to$ C$_2^+$ + H}
\end{equation}
and involves two important molecular species:
methylidyne cation, which is one of the firstly discovered (1941) molecules in
the diffuse ISM \cite{douglas41_381} and whose abundance throughout the
interstellar space still waits for an explanation (see on this
Ref. \cite{myers15_2747} and references therein),
and the dicarbon cation C$_2^+$, which was detected by the mass-spectroscopic
sampling in comets Halley \cite{krankowsky86_326} and Giacobini--Zinner
\cite{coplan87_39} and is incorporated in ion-molecule reactions for the
production of hydrocarbons in interstellar clouds \cite{winnewisser81_39}.

As shall be detailed further on, Reaction \ref{eq:reaction} is exoergic by
about 1.6 eV and involves the barrierless formation of an intermediate
C$_2$H$^+$ complex lying energetically lower than reactants by more than 6.7 eV.
As is known, complex-forming reactions are difficult to characterize
quantum mechanically \cite{guo12_1}.
This is mainly because of the large phase space supported by the potential
well(s), requiring a large basis or grid for converged quantum results.
In addition, because of the attractive potential in the entrance or exit
channel, a large number of partial waves is also needed.
On the other side, experimental studies on ion-radical collisions such as
Reaction \ref{eq:reaction} are also challenging due to the difficulties
associated with making kinetic measurements on processes where both species
are inherently unstable \cite{smith11_29}.
As a consequence, the only available estimates for the thermal rate coefficient 
of this reaction have been for a long time those \cite{woon09_273,wakelam10_13}
based on the simple Langevin capture model (LCM)
\cite{langevin905_245,gioumousis58_294}
and only very recently estimates obtained with quasi-classical trajectory (QCT)
calculations on an \emph{ad hoc} computed potential energy surface have been made available
\cite{pacifici16_5125,rampino16_2368}.
However,  it is well-know that QCT has issues with zero-point energy (ZPE) leakage which can be expected to amplify in the case of very deep well~\cite{rpmd11,qct1,qct2}. Therefore the reliability of the QCT results for the title reaction must be considered with caution.  

Recently, an alternative approach for calculating thermal rate coefficients based on the classical isomorphism~\cite{chandlerwolynes81} between quantum system and its classical ring-polymeric replica (harmonically coupled classical copies of the original system in the form of a necklace) has been proposed~\cite{rpmd0} which is immune to many issues of QCT (as well as transition state rate theory (TST), though this aspect is more relevant to chemical reactions with an activation barrier). The method is called ring polymer molecular dynamics  (RPMD) and approximates real time quantum dynamics by purely classical molecular dynamics of the ring polymer beads. While being purely classical molecular dynamics but in extended phase space, RPMD treats accurately and conserves in its real-time dynamics the quantum Boltzmann distribution and is rigorously independent of the dividing surface used to separate reactant(s) from product(s),  which is particularly challenging to define when the reaction proceeds through a deep well. It also possesses numerous additional features that make this method very attractive for calculating thermal rate coefficients as observed during a comprehensive method assessment on various gas phase atom-diatom and polyatomic chemical reactions \cite{rpmd1,rpmd2,rpmd3,rpmd4,rpmd5,rpmd6,rpmd7,rpmd8,rpmd9,rpmd10,rpmd11,rpmd12,rpmd13,rpmd14,rpmd15,rpmd16,rpmd17,rpmd18,rpmd_ins1,rpmd_ins2,rpmd_ins3} and outlined in the recent review of this method and its practical applications~\cite{rpmdreview}. It has been demonstrated that RPMD is  accurate for prototype atom-diatom insertion chemical reactions~\cite{rpmd_ins1,rpmd_ins2,rpmd_ins3} with deviations from the rigorous quantum dynamics results  close to the convergence error (not exceeding $\sim $ 15 \%). In particular, RPMD provided very accurate estimates of the rate coefficients for the O($^1$D) + H$_2$ reaction, which, similar to the title reaction,  exhibits a deep well ($\sim $ 7.29 eV) and is exoergic ($\sim $ 1.88 eV),~\cite{rpmd_ins1} and for the C($^1$D) + H$_2$ reaction at very low temperatures of astrochemical interest (H-transfer down to 50 K)~\cite{rpmd_ins3}.  
Accurate, consistent and predictable behavior of RPMD distinguishes it from all conventional methods used to calculate thermal rate coefficients~\cite{rpmdreview}. 

Inspired by the previous success of the RPMD rate theory, in particular for barrierless reactions, we carried out the RPMD simulations of the title reaction at temperatures of astrophysical interest (20-300 K) and present them in this paper. 
The remaining sections of the paper are organized as follows.
In Section \ref{sec:method} details on the potential energy surface and on
ring polymer molecular dynamics are given.
In Section \ref{sec:results} results are presented.
In Section \ref{sec:concl} some conclusions are drawn and perspectives for
future work are outlined.
%[Paragraph on ring polymer molecular dynamics (RPMD) \cite{habershon13_387}.
%References for complex-forming reactions \cite{hickson15_4194,li14_700,suleimanov141_244103}.]
%\\
%\ldots \\
%\ldots \\
%\ldots
%RPMD rate coefficients are (i)
%exact in the high-temperature limit, (ii) reliable at intermediate
%temperatures, (iii) more accurate than other approximate
%methods in the deep quantum tunnelling regime and close to
%the exact quantum results, and (iv) able to capture ZPE effects
%perfectly.]
%[(near) barrierless
%formation of long-lived complexes in deep potential wells
%C($^1$D) + H$_2$ reaction \cite{hickson15_4194}.
%Complex-forming X $=$ N, O \cite{li14_700} and X $=$ C($^1$D), S($^1$D) \cite{suleimanov141_244103}.

%[In this paper \ldots]
%\\
%\ldots \\
%\ldots \\
%\ldots

\section{Methodology and computational details}\label{sec:method}

\subsection{Potential energy surface}\label{sec:method:pes}

The PES used for the RPMD calculations carried out in this work has already
been described in Refs. \cite{pacifici16_5125} and \cite{rampino16_2368}.
Here we briefly recall the methodology used to assemble it and describe its
main features relevant to the dynamics.

The PES, in the form of a Fortran routine for use in dynamics programs and
available upon request to the authors,
\bibnote{The PES is available upon request to S.R. (\url{info@srampino.com})}
was obtained by fitting the well known Aguado--Paniagua
\cite{aguado92_1265,aguado98_259} functional form to a set of 775 three-body
and 20 two-body electronic energies obtained by second-order multi-reference
perturbation theory (MRPT) in the `partially contracted' PC-NEVPT2 scheme
\cite{angeli02_9138,angeli07_743}.
Configuration-space sampling (ie., the choice of the geometries at which the
\emph{ab initio} calculations were run) was performed according to the
space-reduced bond-order (SRBO) approach recently published by one of us (S.R.)
\cite{rampino16_4683}.
In the SRBO scheme, use is made of opportunely defined diatomic bond-order
(BO) variables
%
%\begin{equation}
%n = e^{-\beta_\mathrm{BO} (r - r_\mathrm{e})}
$n = \exp[-\beta (r - r_\mathrm{e})]$
%\end{equation}
%
(with $r$ being the diatom internuclear distance and $r_\mathrm{e}$ its
equilibrium value)
where $\beta$ is relaxed so as to reach a desired ratio $f$ between the
sampled attractive ($0 < n<1$) and repulsive ($1< n <e^{\beta r_\mathrm{e}}$)
regions of the diatom configuration space (see also Ref.
\cite{rampino12_1818}).
A proper tuning of $f$ and the adoption of regular grids in SRBO variables
allows for a wise, process-oriented selection of geometries having built-in a
force-based metric and thus providing a small, most informative set of
electronic energies.
\bibnote{A Fortran computer program for constructing SRBO grids is available
at \url{http://www.srampino.com/code.html#Pestk} or upon request to S.R.}

\begin{figure}
\includegraphics[width=0.75\textwidth]{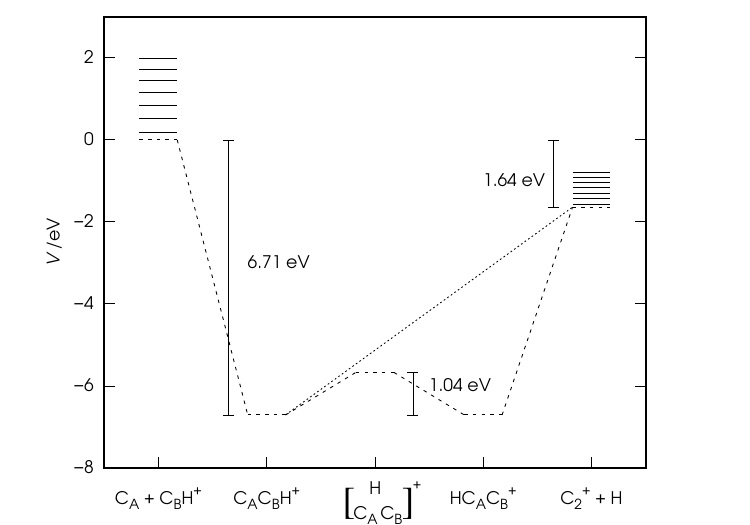}
\caption{%
Energy profile for Reaction \ref{eq:reaction}. 
The vibrational structure
($v = $ 0--6, $j = $ 0)
of CH$^+$ and C$_2^+$ is also shown at the reactant (left) and product (right) asymptote, respectively. 
The reader is referred to Section \ref{sec:method:pes} for a discussion.
}\label{figs1}
\end{figure}

The main features of the resulting PES concerning the title reaction are schematized in Fig. \ref{figs1} and can be
summarized as follows.
The reaction is exoergic by 1.64 eV and proceeds through the barrierless
formation of a C$_2$H$^+$ intermediate.
The energetically favoured reaction path is the collinear one with the reactant
carbon C$_\mathrm{A}$ atom approaching C$_\mathrm{B}$H$^+$ from the carbon side
(where labels A and B have been adopted to distinguish between the two carbon atoms)
and leading to
formation of a linear C$_\mathrm{A}$C$_\mathrm{B}$H$^+$ triatomic sitting at the bottom
of a potential well which is
as deep as 6.71 eV measured from the bottom of the reactants channel.
At this point the system can either dissociate into products H and C$_2^+$
or explore a second, identical potential well due to rotation of the hydrogen atom
about the carbon-carbon bond.
This last path involves overcoming a rotational
barrier of 1.04 eV before forming the linear triatomic HC$_\mathrm{A}$C$_\mathrm{B}$$^+$ and further proceed to products.
The reader is referred to Ref. \cite{pacifici16_5125} for a more detailed
discussion on these alternative reaction paths and their
effects
on the QCT reaction dynamics.
In Figure \ref{figs1}, the vibrational structure (first seven vibrational levels) of diatomics CH$^+$ and C$_2^+$ is also reported.
When including zero-point energies (0.18 and 0.07 eV for the reactant and product diatom, respectively) the exoergicity of the title reaction amounts to 1.75 eV.

\subsection{Ring polymer molecular dynamics}\label{sec:method:rpmd}
The RPMD calculations were performed using the RPMDrate code developed by one of us (Y.V.S.)~\cite{rpmd6}.  
The computational procedure for calculating RPMD rate coefficients is well documented in the RPMDrate manual~\cite{rpmd6} and in the recent review of the RPMD rate theory and its practical applications.\cite{rpmdreview} The reader is referred to the corresponding reference for more detail. 

\begin{figure}
\includegraphics[width=0.75\textwidth]{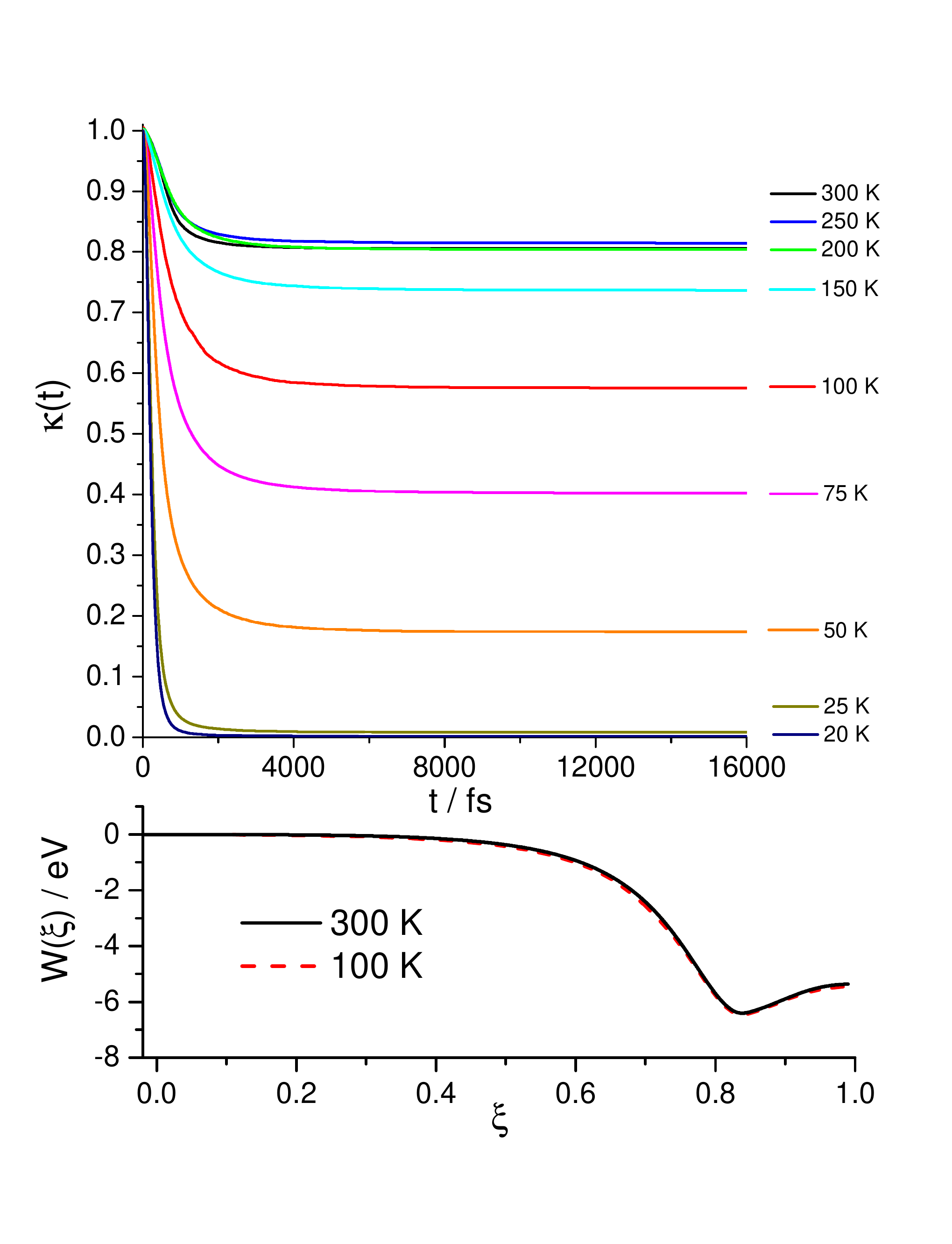}
\caption{Ring polymer transmission coefficient (upper panel) at $T$ = 20-300 K and centroid potential of mean force (lower panel) at $T$ = 100 and 300 K. (Only two temperatures were included in the lower panel due to the visual indistinguishability of the results). 
%The plateau values of transmission coefficients are given in Table \ref{tab:rpmd}.
}\label{figs2}
\end{figure}

The ring polymer transmission coefficients at all temperatures of the present study (20-300 K) and potentials of mean force (free energy) at two representative temperatures (100 and 300 K) are depicted in Figure~\ref{figs2}. 
The simulation parameters are summarized in Table~\ref{tabs1}.
\begin{table*}
\scriptsize
\caption{ Input parameters for the RPMD calculations on the title reaction. 
The explanation of the format of the input file can be found in the RPMDrate code manual 
(http://rpmdrate.cyi.ac.cy).}
\label{tabs1}
\begin{center}
\begin{tabular}{lll} 
\hline
\hline
Parameter & Reaction & Explanation \\
\cline{2-3}
 & C + CH$^+$ $\to$ C$_2^+$ + H    & \\
\hline
\multicolumn{3}{l}{Command line parameters} \\ 
\hline
\ttfamily{Temp}         & 20; 25; 50; 75; 100; 150; 200; 250; 300      &       Temperature (K)  \\

\ttfamily{Nbeads}       & 128 & Number of beads \\	
\hline
\multicolumn{3}{l}{Dividing surface parameters} \\ 
\hline
$R_\infty $  & 15 $a_0$   &    Dividing surface parameter (distance) \\
$N_{\rm bonds}$       & 1    & Number of forming and breaking bonds \\
$N_{\rm channel}$     &1      & Number of equivalent product channels \\
C(CH$^+$) & (-1.09 $\angstrom $, 0.00 $\angstrom $, 0.00 $\angstrom $) &  Cartesian coordinates (x, y, z) \\
H(CH$^+$) & (0.00 $\angstrom $, 0.00 $\angstrom $, 0.00 $\angstrom $) & of the intermediate geometry \\ 
C               & (1.27 $\angstrom $, 0.00 $\angstrom $, 0.00 $\angstrom $) & \\ 
\hline
\ttfamily{Thermostat}  & 'Andersen' &  Thermostat option \\
\hline
\multicolumn{3}{l}{Biased sampling parameters}\\ 
\hline
$N_{\rm windows}$  & 7(111)$^a$           & Number of windows \\
$\xi _1$            & -0.05      & Center of the first window \\
$d\xi $              & 0.01         & Window spacing step      \\ 
$\xi _N$             & 0.01 (1.05)       & Center of the last window \\
$dt$  & 0.0001 &   Time step (ps) \\
$k_i$  &  2.72  & Umbrella force constant ((T/K) eV) \\
$N_{\rm trajectory}$ & 200   & Number of trajectories \\ 
$t_{\rm equilibration}$ & 20      & Equilibration period (ps)  \\
$t_{\rm sampling}$ &  100     & Sampling period in each trajectory (ps) \\
$N_i$   & $2\times 10^8$ &   Total number of sampling points \\
\hline
\multicolumn{3}{l}{Potential of mean force calculation} \\ 
\hline
$\xi _0$            & 0.00     & Start of umbrella integration  \\
$\xi ^{\ddagger}$             & 0.0031$^{a,b}$   & End of umbrella integration \\

$N_{\rm bins}$             & 5000  &  Number of bins \\
\hline
\multicolumn{3}{l}{Recrossing factor calculation} \\ 
\hline
$dt$  & 0.0001 & Time step (ps) \\
$t_{\rm equilibration}$  & 20  &  Equilibration period (ps) in the constrained (parent)\\ 
 & &    trajectory \\
$N_{\rm totalchild}$  & 300000   & Total number of unconstrained (child) trajectories  \\
$t_{\rm childsampling}$  & 2    & Sampling increment along the parent trajectory  (ps)  \\
$N_{\rm child}$  & 50  &   Number of child trajectories per one   \\
  & &   initially constrained configuration \\
$t_{\rm child}$  & 16    &Length of child trajectories (ps)  \\
\hline
\hline
\end{tabular}\\
\end{center}
$^a$ Complete umbrella integration with 111 windows up to $\xi _N$ = 1.05 was performed only at 100 and 300 K for Fig.~\ref{figs2}; \\
$^b$ Set fixed for all temperatures. For 100 and 300 K the potential of mean force was additionally reconstructed up to $\xi _N$ = 1.05 for Fig.~\ref{figs2}.  \\
\end{table*}
We found that 128 ring polymer beads were sufficient to converge the RPMD rate coefficients at all temperatures. The RPMD treats the atoms as distinguishable but because the title reaction is the C-to-C transfer the nuclear spin statistics is expected to play negligible role even at the lowest temperature  (T = 20 K). 
The remaining simulation parameters are similar to those used in the previous RPMD studies of insertion chemical reactions~\cite{rpmd_ins1,rpmd_ins2,rpmd_ins3} but with two distinctions. First, umbrella integration is terminated near reactants as the title reaction exhibits very deep well at the  free energy profile (see lower panel of Fig.~\ref{figs2}) and further propagation may only lead to enhancing the ring polymer recrossing dynamics which, in its turn, leads to poorer convergence of the final RPMD rate coefficients. (For the title reaction, we haven't observed any tiny free-energy barrier before the entrance into the C$_2$H$^+$ well.) Second, the ring polymer recrossing dynamics is propagated up to 16 ps ($t_{\rm child}$ in Table~\ref{tabs1}) which, to our knowledge is the longest real-time dynamics ever taken into account in the RPMD studies of chemical reactions. This is again due to the presence of a very deep well along the reaction coordinate. 
Figure~\ref{figs2} shows that the free energy profile is practically temperature-independent but the transmission coefficients significantly decrease with decreasing the temperature. Note that at very low temperatures the recrossing dynamics is enhanced for the title reaction leading to very small plateau values which were challenging to converge. In future studies of such chemical reactions at low temperatures, RPMD can be coupled with the parallel replica dynamics approach~\cite{voter2016}.

%\subsection{Computational details}\label{sec:method:details}

\section{Results}\label{sec:results}

%\subsection{Thermal rate coefficients}

\begin{figure}
\includegraphics[width=0.75\textwidth]{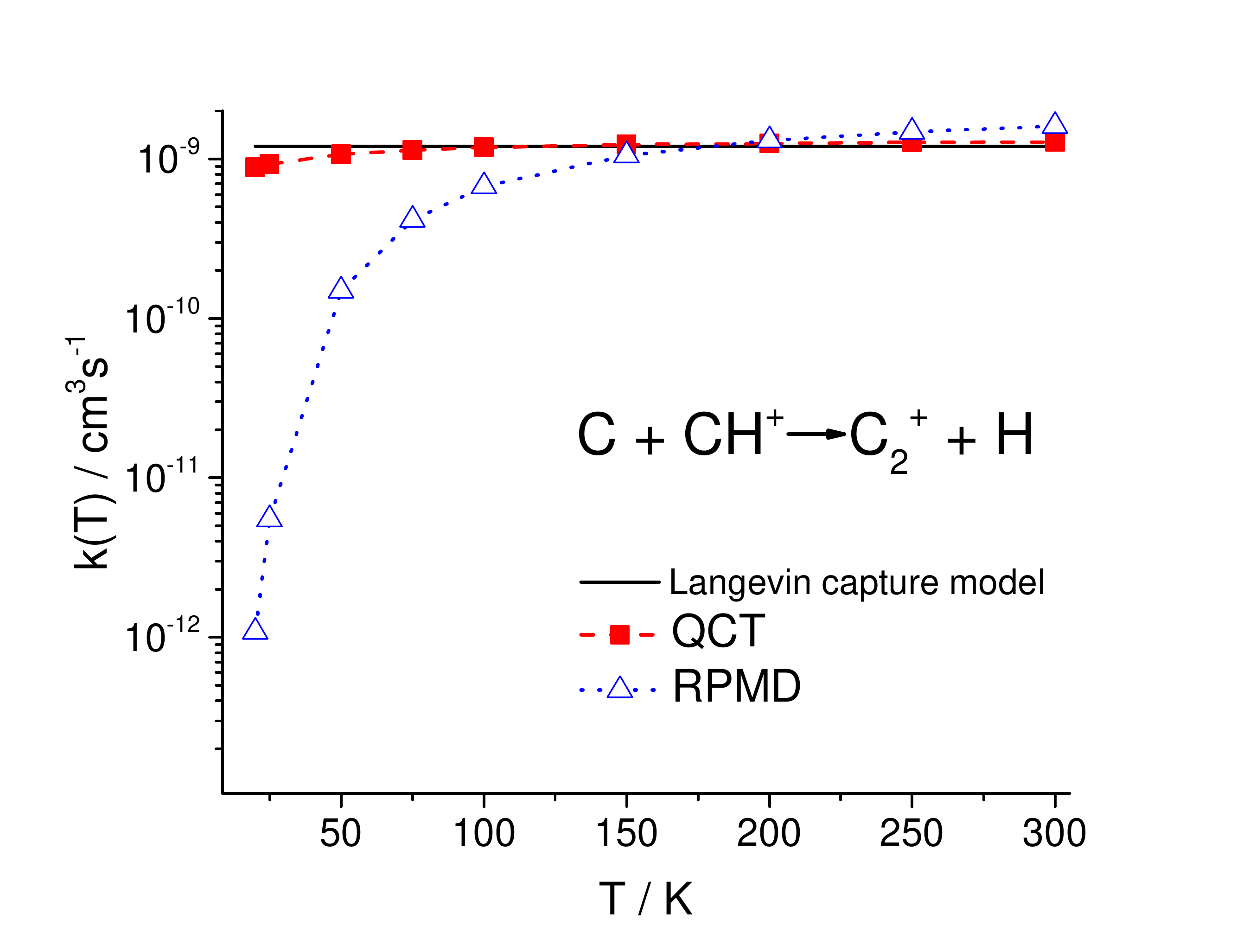}
\caption{Langevin capture model, QCT and RPMD thermal rate coefficients plotted as a function of temperature $T$ (QCT and RPMD data are given in Table \ref{tabs2}).}\label{figs3}
\end{figure}

The RPMD thermal rate coefficients are compared with the previous QCT ones in Table~\ref{tabs2} and are plotted in Figure~\ref{figs3}, which also includes the LCM estimate. 
\begin{table}
\caption{%
QCT and RPMD values of $k(T)$ for the considered set of temperatures.$^a$
}\label{tabs2}
\begin{center}
\begin{tabular}{rrr}
\hline
\hline
$T$/K & \multicolumn{2}{c}{$k(T)$/cm$^3$s$^{-1}$} \\
\cline{2-3} \
  &
    \multicolumn{1}{c}{QCT} &
%    \multicolumn{1}{c}{QTST} &
    \multicolumn{1}{c}{RPMD} \\
\hline
%
%15   & \rndd{8.1999} $\times 10^{-10}$ & 
%5.63 $\times 10^{-10}$ & 
%1.38 $\times 10^{-13}$ \\
20   & \rndd{8.8302} $\times 10^{-10}$ & 
%6.15 $\times 10^{-10}$ & 
1.08 $\times 10^{-12}$ \\
25   & \rndd{9.31875} $\times 10^{-10}$ & 
%6.55 $\times 10^{-10}$ & 
5.48 $\times 10^{-12}$ \\
50   & \rndd{1.0714} $\times 10^{-09}$ & 
%8.60 $\times 10^{-10}$ & 
1.49 $\times 10^{-10}$ \\
75   & \rndd{1.1405} $\times 10^{-09}$ & 
%1.03 $\times 10^{-09}$ & 
4.15 $\times 10^{-10}$ \\
100  & \rndd{1.1811} $\times 10^{-09}$ & 
%1.18 $\times 10^{-09}$ & 
6.72 $\times 10^{-10}$ \\
150  & \rndd{1.2305} $\times 10^{-09}$ & 
%1.42 $\times 10^{-09}$ & 
1.05 $\times 10^{-09}$ \\
200  & \rndd{1.2520} $\times 10^{-09}$ & 
%1.64 $\times 10^{-09}$ & 
1.31 $\times 10^{-09}$ \\
250  & \rndd{1.2654} $\times 10^{-09}$ & 
%1.82 $\times 10^{-09}$ & 
1.48 $\times 10^{-09}$ \\
300  & \rndd{1.2751} $\times 10^{-09}$ & 
%1.99 $\times 10^{-09}$ & 
1.61 $\times 10^{-09}$ \\
\hline
\hline
\end{tabular} \\
\end{center}
$^a$QCT values taken from
Ref.\cite{rampino16_2368}. Note that QCT values of $k(T)$ at 25 and 75 K were obtained by linear interpolation of the available neighbouring values (24 and 26 K and 70 and 80 K, respectively).%
\end{table}
Figure~\ref{figs3} shows
that the QCT (red squares) and RPMD (blue triangles) thermal rate coefficients are lower than the Langevin estimate (black solid line) at the lowest temperatures, and that both tend to increase with increasing $T$ getting to cross the Langevin value in the temperature interval between 100 and 200 K.
However, despite this qualitative similarity in the temperature dependence, QCT and RPMD results differ substantially, especially as the temperature goes down to 20 K.
In fact, whereas the QCT and RPMD curves cross each other at about 200 K
and at the highest considered temperature (300 K) both the QCT and RPMD results agree with the LCM estimate
within
approximately
25 \%,
at 20 K the discrepancy between QCT and RPMD results attains several orders of magnitude.
It is worth recalling here that QCT does not take into account the change in ZPE
along the reaction coordinate and that the
problem with ZPE can be particularly significant in the case of barrierless reactions as the title one with very deep well and complex profile of the minimum energy path as depicted in Figure~\ref{figs1}. The QCT trajectories exiting the potential well may in fact violate the ZPE and lead to substantial erroneous increase of the rate coefficients~\cite{rpmd11,qct1,qct2}.

%\subsection{Parametrized formulations of $k(T)$}

As mentioned in the Introduction,
thermal rate coefficients
for use in astrochemical kinetic networks are made
available through dedicated databases such as the already quoted OSU, KIDA and UDfA.
Thermal rate coefficients are therein provided 
 in the form of a set of parameters expressing their temperature dependence via popular parametrized formulations.
Rather than the original
Arrhenius equation
\begin{equation}
k(T) = A e^{-\frac{E_\mathrm{a}}{RT}} %\;,
\label{eq:ar}
\end{equation}
(with $A$ being the pre-exponential factor, $E_\mathrm{a}$ the activation energy, and $R$ the gas constant)
the 
Arrhenius--Kooij formula \cite{kooij893_155} (also known to chemists
as modified Arrhenius equation \cite{laidler96_149} and allowing for a temperature dependence of the pre-exponential factor)
\begin{equation}
%k(T) = A (T/300)^d e^{-\epsilon/RT} \;,
k(T) = \alpha (T/300)^\beta e^{-\gamma/T} %\;,
%\ln k(T) = \ln \alpha + \beta \ln (T/300) -\gamma/T
\label{eq:ak}
\end{equation}
is adopted in astrochemical kinetic networks.
In the already quoted Ref.
\cite{rampino16_2368},
we also investigated the suitability of
the so-called `deformed Arrhenius' equation recently proposed by Aquilanti and Mundim\cite{aquilanti10_209,aquilanti12_186}
\begin{equation}
k(T) = A \left[ 1 - d \frac{\epsilon}{RT}\right]^{\frac{1}{d}} %\;,
\label{eq:da}
\end{equation}
to account for deviations from the Arrhenius behaviour.
We found therein that indeed the temperature dependence of the QCT thermal rate coefficients for the same reaction analyzed in this paper better conforms to the `Aquilanti--Mundim' law.

\begin{figure}
\includegraphics[width=0.75\textwidth]{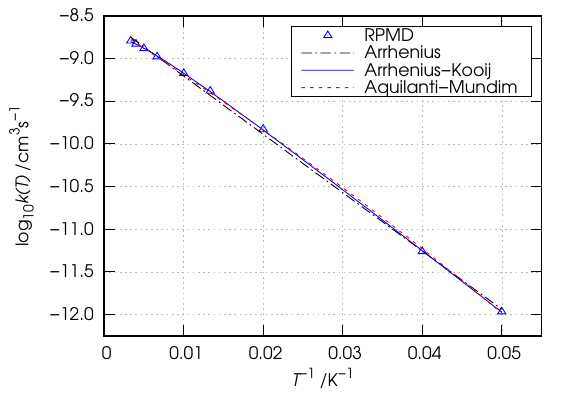}
\caption{Arrhenius plot of the RPMD thermal rate coefficient (blue triangles) and of the Arrhenius (dashed-dotted black line), Arrhenius--Kooij (solid blue line) and Aquilanti--Mundim (dashed red line) best-fitting curves. 
}\label{figs4}
\end{figure}

For the sake of comparison and to the purpose of providing more reliable estimates for the rates of the title process to be included in astrochemical kinetic networks,
we
performed the non-linear fits of Eqs. \ref{eq:ar}-\ref{eq:da} to the computed RPMD thermal rate coefficient in the temperature range 20-300 K to determine the related best-fitting parameters.
Results of the fitting procedures are summarized in Table \ref{tabs3}, including values of $\chi^2$ and correlation coefficients.
The Arrhenius, Arrhenius--Kooij and Aquilanti--Mundim best-fitting curves are also shown in Fig. \ref{figs4} as dashed-dotted black line, solid blue line and dashed red line, respectively, for a comparison with the computed RPMD values (blue triangles).
According the present RPMD calculations, the thermal rate coefficients for the title reaction display substantially Arrhenius behaviour with the Arrhenius--Kooij and Aquilanti--Mundim best-fitting curves excellently reproducing the computed data (correlation coefficient $>0.995$).
\begin{table}
\centering
\caption{%
Results of the non-linear fit of the Arrhenius,
Arrhenius--Kooij
%(Eq. \ref{eq:ak})
and Aquilanti--Mundim
%(Eq. \ref{eq:da})
equations to the computed
RPMD thermal rate coefficients in the temperature range 20-300 K.$^a$
}
\label{tabs3}
\begin{center}
\begin{tabular}{ccccc}
  \multicolumn{5}{c}{} \\
  \multicolumn{5}{c}{Arrhenius (Eq. \ref{eq:ar})} \\
\hline
 $A$ (cm$^3$s$^{-1}$) & & $E_\mathrm{a}/R$ (K) & $\chi^2$ & corr. coeff. \\
 $\rndd{3.017}\times 10^{-9}$ & &  $\rndd{157.347}$ & $\rndq{0.0622274}$ & $\rndq{0.9788657}$ \\
  \multicolumn{5}{c}{} \\
  \multicolumn{5}{c}{Arrhenius--Kooij (Eq. \ref{eq:ak})} \\
\hline
 $\alpha$ (cm$^3$s$^{-1}$) & $\beta$ & $\gamma$ (K) & $\chi^2$ & corr. coeff. \\
 $\rndd{2.8055}\times 10^{-9}$ & $\rndd{-0.2699}$ & $\rndd{172.103}$ & $\rndq{0.00235148}$ & $\rndq{0.9992095}$ \\
  \multicolumn{5}{c}{} \\
  \multicolumn{5}{c}{Aquilanti--Mundim (Eq. \ref{eq:da})} \\
\hline
 $A$ (cm$^3$s$^{-1}$) & $d$ & $\epsilon/R$ (K) & $\chi^2$ & corr. coeff. \\
 $\rndd{2.653}\times 10^{-9}$ & $\rndd{0.0336}$ & $\rndd{137.975}$ & $\rndq{0.0137015}$  & $\rndq{0.99538505}$ \\
\end{tabular} \\
\end{center}
$^a$Note that the non-linear fit was performed using the natural logarithm of both the RPMD data and Eqs. \ref{eq:ar}-\ref{eq:da}.
\end{table}

\section{Conclusions and perspectives}\label{sec:concl}

In this paper we use ring polymer molecular dynamics (RPMD) to compute thermal
rate coefficients $k(T)$ for reaction C + CH$^+$ $\to$ C$_2^+$ + H in the temperature
range 20-300 K on a recently published potential energy surface.
Results are compared with previous estimates based on Langevin capture model (LCM) and quasi-classical trajectory (QCT) calculations.
Non-linear fits of the computed RPMD data are also carried out using several
parametrized formulations of $k(T)$ and the related best-fitting parameters
are given for inclusion in astrochemical databases.
In the absence of rigorous quantum dynamics or experimental results,
the present RPMD calculations are expected to provide the most accurate available estimates for the thermal rates of the title reaction to be used in astrochemical kinetic networks.

In a previous work of ours~\cite{rampino16_2368}, it has been shown that QCT calculations of the rate coefficients for this reaction lead to a deviation from the LCM value as the temperature is decreased from the room one to the astrochemical diapason.
The QCT thermal rate coefficient at 10 K resulted, in fact, lower than the LCM estimate by a factor of two.
However, the present results suggest that even the QCT calculations substantially overestimate the rate of C$_2^+$ formation (and, consequently, CH$^+$ consumption) at low temperatures leading to an error of several orders of magnitude at 20 K.
Accordingly, the present results also partially address the already mentioned issue of the unexplained observed abundance of methylidyne cation throughout the interstellar medium.
Though in this respect
reactions of CH$^+$ with atomic and molecular hydrogen are undoubtedly more decisive than the reaction with atomic carbon, the present results show that the destruction route of CH$^+$ is erroneously enhanced in kinetic models using either LCM or QCT estimates for the process considered in this paper.

The title reaction exhibits a very complex energy profile along the reaction coordinate as shown in Figure~\ref{figs1}.
As is known, proper treatment of the zero-point energy along the reaction coordinate is the Achilles' heel of many approximations including QCT.
RPMD is immune to this issue as well as it possesses several other advantages over conventional approaches for calculating rate coefficients~\cite{rpmdreview}, especially for astrochemical elementary reactions (with either deep well~\cite{rpmd_ins3} or barrier~\cite{rpmd9} along the reaction coordinate), where quantum mechanical effects of nuclear motions play a crucial role. 
We hope therefore that RPMD will find wide application in improving astrochemical kinetic databases in the future.

%\appendix
%\section{Appendix: temporary stuff}
%\subsection{Bibliographical annotations}

%%%%%%%%%%%%%%%%%%%%%%%%%%%%%%%%%%%%%%%%%%%%%%%%%%%%%%%%%%%%%%%%%%%%%
%% The "Acknowledgement" section can be given in all manuscript
%% classes.  This should be given within the "acknowledgement"
%% environment, which will make the correct section or running title.
%%%%%%%%%%%%%%%%%%%%%%%%%%%%%%%%%%%%%%%%%%%%%%%%%%%%%%%%%%%%%%%%%%%%%
\begin{acknowledgement}

S.R. thanks the European Research Council for funding through the European Union's Seventh Framework Programme (FP/2007-2013) / ERC Grant Agreement n. [320951] and Prof. V. Barone for useful discussion.
Y.V.S. thanks the European Regional Development Fund and the Republic of Cyprus for support through the Research Promotion Foundation (Project Cy-Tera NEA ${\rm Y\Pi O\Delta OMH}$ / ${\rm \Sigma TPATH}$/0308/31).

\end{acknowledgement}

%%%%%%%%%%%%%%%%%%%%%%%%%%%%%%%%%%%%%%%%%%%%%%%%%%%%%%%%%%%%%%%%%%%%%
%% The same is true for Supporting Information, which should use the
%% suppinfo environment.
%%%%%%%%%%%%%%%%%%%%%%%%%%%%%%%%%%%%%%%%%%%%%%%%%%%%%%%%%%%%%%%%%%%%%
%\begin{suppinfo}
%
%% srampino
%[No SI at the moment.]
%% end srampino
%
%\end{suppinfo}

%%%%%%%%%%%%%%%%%%%%%%%%%%%%%%%%%%%%%%%%%%%%%%%%%%%%%%%%%%%%%%%%%%%%%
%% The appropriate \bibliography command should be placed here.
%% Notice that the class file automatically sets \bibliographystyle
%% and also names the section correctly.
%%%%%%%%%%%%%%%%%%%%%%%%%%%%%%%%%%%%%%%%%%%%%%%%%%%%%%%%%%%%%%%%%%%%%
\bibliography{./jpca16}

\end{document}